\documentclass[preprint]{elsarticle}
\usepackage{graphicx,url,hyperref,microtype,subcaption,graphbox,amsmath}
\usepackage{latexsym,courier,upgreek,xcolor,rotating,ulem}
\usepackage[onehalfspacing]{setspace}
\hypersetup{colorlinks,citecolor=blue,linkcolor=red,urlcolor=green}
\usepackage[displaymath,mathlines]{lineno}
\usepackage[margin=2cm]{geometry}
\modulolinenumbers[5]
\biboptions{numbers,sort&compress}

\begin{document}

\begin{frontmatter}
  
  \title{From nucleation to percolation: the effect of system size when disorder and stress localization compete}
  
  \author{Subhadeep Roy}
  \ead{subhadeeproy03@gmail.com}
  
  \address{PoreLab, Department of Physics, Norwegian University
    of Science and Technology, N-7491 Trondheim, Norway.}

  \begin{abstract}
   A phase diagram for a one-dimensional fiber bundle model is constructed with a continuous variation in two parameters guiding the dynamics of the model: strength of disorder and range of stress relaxation. When the range of stress relaxation is very low, the stress concentration plays a prominent role and the failure process is nucleating where a single crack propagates from a particular nucleus with a very high spatial correlation unless the disorder strength is high. On the other hand, a high range of stress relaxation represents the mean-field limit of the model where the failure events are random in space. At an intermediate disorder strength and stress release range, when these two parameters compete, the failure process shows avalanches and precursor activities. As the size of the bundle is increased, it favors a nucleating failure. In the thermodynamic limit, we only observe a nucleating failure unless either the disorder strength is extremely high or the stress release range is high enough so that the model is in the mean-field limit. A complex phase diagram on the plane of disorder strength, stress release range, and system size is presented showing different failure modes - (I) nucleation, (II) avalanche, and (III) percolation, depending on the spatial correlation observed during the failure process.
  \end{abstract}

  \date{\today}
  
  
\end{frontmatter}


\makeatletter
\def\ps@pprintTitle{%
 \let\@oddhead\@empty
 \let\@evenhead\@empty
 \def\@oddfoot{}%
 \let\@evenfoot\@oddfoot}
\makeatother

\section{Introduction}

It is nearly a century since Alan Arnold Griffith developed his energy criterion for the fracture propagation of cracks in near-continuous solids \cite{citation1,citation2}. His celebrated work has revolutionized the world of materials science. Griffith considered a single sharp crack in an otherwise homogeneous elastic medium. In Griffith's theory, the crack propagation is considered as an equilibrium problem where the balance between two energies: reduction of strain energy, and increment in surface energy is measured during the crack propagation. He found that the critical stress $\sigma_c$ to cause a crack of length $l$, to extend is $\sigma_c=(2Yg/\pi l)^{1/2}$ \cite{knott73}, where $Y$ is Young's modulus and $g$ is the surface energy per unit area. However, this is an idealized case that requires a pre-existing crack or notch in a homogeneous medium to concentrate the applied stress. In general, the initiation of a fracture in a solid is a much more complex process. Most engineering materials are far from homogeneous, there will always be a distribution of dislocations, flaws, and other heterogeneities present. The nucleation and propagation of a crack in heterogeneous systems are not understood because of the complexities of the stress singularities at the crack tip \cite{Phoenix}. As the applied stress is increased, micro-cracks are likely to occur randomly on the heterogeneity and are uncorrelated. As the density of micro-cracks increases, the stress fields of the micro-cracks interact and the micro-cracks become correlated. The micro-cracks eventually may coalesce to form a through-going fracture. This irreversible process is a part of damage mechanics and is an integral part of the nucleation and propagation of fracture in heterogeneous environments.

During the failure process of a disordered system, a complex interplay is observed between quenched heterogeneities and local stress concentration. The former one leads to non-localized damage mechanics while the latter favors the formation of localized cracks. As a consequence of this interplay, we observe system size dependence of nominal stress distribution \cite{hp78,bazabt04,anz06}, scale-free avalanche size statistics \cite{hh92,ppvac94,ggbc97,sta02}, self-affine crack morphology \cite{bb11}, etc. In the limit of infinitesimal disorder strength, the crack grows within a disordered system in a nucleating manner \cite{dbl86,dbl87,cb97,msnasz12}. On the other hand, when disorder strength is infinitely high, the effect of local stress concentration becomes irrelevant and the failure process is random in space \cite{rahg88,hs03} like percolation. The reason behind the damage mechanics at high disorder is the heterogeneities in a material that create energy barriers which act as a resistance on the way of crack propagation and ultimately arrest the crack motion: a phenomenon known as lattice trapping or intrinsic crack resistance \cite{thr71,cm90,rice92,pg00,bh03,mcc05}. At intermediate disorder, the situation is more interesting, where the failure process takes place through a number of avalanches showing scale-free distributions of energies emitted during the avalanches \cite{ppvac94} and mean-field exponents \cite{hh94,zrsv97,zns05,zns05a}. 

Tuning the strength of disorder is not achievable easily in experiments. Though through heat treatment one can tune the length scale of disorder in phase-separated glasses \cite{dlv08}. Earlier experiments also studied the role of a varying disorder strength during pattern formation in random spring network \cite{mm07,rd96,pmbr20}, the study of roughness of a fracture surface \cite{s06}, the transition from nucleation to damage mechanics in random fuse network \cite{szs13,mohaha12}, etc. Linear elastic fracture mechanics, on the other hand, predicts the load distribution around an Inglis crack \cite{srvm95,ref97} to be $1/r^2$-type where $r$ is the distance from the crack tip. However, this form for relaxation of local stress can be affected by many parameters like correlation among defects \cite{phss91} and effect of the limited size of the sample \cite{xh00}. This in turn can change the dynamics of crack propagation. We will explore here the effect all there important parameters: disorder strength, stress release range and sample size in detail.  

In this article, we study the spatial correlation during failure process of a fiber bundle model (FBM) \cite{hansenBook,cbp21}. Fiber bundle model is very effective yet arguably the most simplest model to understand failure process of heterogeneous materials. The effect of a variable stress release range has already been observed in the context of this model \cite{brr15,hmkh02}. With a very low stress stress release range, the failure process is observed to be nucleating and at the same time, the failure abruptness is affected by the size of the bundle \cite{brr15,r17}. The spatial correlation, in this limit, decreases as the disorder strength is increased \cite{srh20}. On the other hand, at a very high stress release range (the mean-field limit), the failure abruptness is not a function of system size and only controlled by the strength of disorder \cite{rr15}. A high thermal fluctuation as well leads to a failure process, random in space, with the same universality class of a site percolation \cite{srh21}. When we combine both the effects of stress release range and disorder strength, the model produces rich relaxation dynamics with different modes of failure $-$ abrupt, non abrupt, nucleating, and random in space \cite{rbr17}. The avalanche statistics as well as the effect of system sizes has also been discussed recently in the context of fiber bundle model \cite{kdk17,kk19}. Moreover, the record statistics in the avalanche statistics and study of elastic energy has been observed to be a vital key to predict an upcoming failure \cite{kpk20,pkh19}. The occurrence of different regimes with increasing disorder strength is also observed in spin systems as well. Recent work in the random field Ising model shows many small avalanches at high disorder and depinning behavior when disorder strength is very weak. At an intermediate critical value of disorder strength, avalanches of all possible sizes are observed \cite{tmjsr19,smpv18}. A study similar to the present one is explored in the context of random fuse network \cite{szs13,mohaha12}. In the former study, the avalanche size distribution as well as the system size scaling of average avalanche size and average crack size. In the latter study, the authors have studied the masses of the largest cluster, the mass of the backbone (the number of burnt bonds forming the chain that effectively disconnects the bottom from the top of the system), and how those masses scale with the system size in order to compare them with percolation. Here we carry out a study, similar to what was observed in the resistor network for a variable stress release range and construct a phase diagram on the plane of disorder, stress release range and system size. We have mainly studied the behavior of crack density to observe the spatial correlation. The crack density is explored earlier by the same author \cite{brr15,rbr17} but a systematic study with system size, especially a systematic study of maximum crack density with the size of the bundle was missing there. Such a study is carried out in the present article and can offer a nice insight into the fracture pattern. 


\section{Description of Fiber Bundle Model}

After its introduction by Pierce in 1926 \cite{Pierce}, the fiber bundle model \cite{hansenBook} has been proven to be very effective in understanding the failure event of a disordered system. Due to this, fiber bundle model is growing popularity among engineering, material science as well as academics. This model is very effective yet arguably the most simple model guided by threshold activated dynamics. 

A conventional fiber bundle model consists of $L$ parallel Hookean fibers in between two plates which are pulled apart creating a stress $\sigma$ on each fiber. The fluctuation among the strength values of individual fibers is the measure of disorder within the model. In the present work, we chose such strength values ($h$) from a power law distribution with a slope $-1$ and span from $10^{-\beta}$ to $10^{\beta}$. 
\begin{equation}\label{eq4}
p(h) \sim \begin{cases}
    h^{-1},  & (10^{-\beta} \le h \le 10^{\beta}) \\
    0.  & ({\rm otherwise})
  \end{cases}
\end{equation}
Here $\beta$ is related to the span of the distribution and a measure of disorder strength. We have chosen such a distribution as such long-tailed power-law distribution \cite{ft28} has already been observed for the distribution of material strength. A certain fiber breaks irreversibly when the stress applied on it exceeds its strength. The stress of that fiber is then redistributed among rest of the model. Here, we have adopted a generalized rule for stress redistribution. If $\sigma_i$ is the stress on the broken fiber $i$, then the stress redistribution on fiber $j$ at a distance $r_{ij}$ from fiber $i$ will be as follows      
\begin{equation}\label{eq2}
\sigma_j \rightarrow \sigma_j + \displaystyle\frac{r_{ij}^{-\gamma}}{Z} \sigma_i
\end{equation}
where $\sigma_j$ is the stress on fiber $j$ and $Z$ is the normalization factor given by
\begin{equation}\label{eq3}
Z = \displaystyle\sum_{i,k} r_{ik}^{-\gamma}
\end{equation}
where $k$ runs overall intact fibers. Two extreme limits of this redistribution rule are global load sharing (GLS) \cite{Pierce} and local load sharing (LLS) \cite{Phoenix,Smith} limit. $\gamma$ has a very low value for the former case and stress of the broken fiber here is redistributed among all surviving fibers in almost same amount. This is also the mean-field limit of the model. In the other extreme limit, $\gamma$ has a very high value and a large amount of the redistributed stress is carried by the neighboring fibers of the broken one only. The effect of stress concentration is most prominent here. Earlier study shows a critical value $\gamma_c$ of the stress release range, for both 1d \cite{brr15} and 2d \cite{hmkh02}, around which the model transit from the mean-field limit to the local load sharing limit. After such redistribution, due to the elevated local stress profile, further fibers may also break starting an avalanche. With such a process, the bundle may break through a single avalanche or comes to a steady state with some fibers broken and some intact. In the later situation, the external stress is increased to break the next weakest fiber starting a new avalanche. Such process goes on until all fibers are broken.           
 

\section{Numerical Results}

A one dimensional fiber bundle model is studied numerically with varying disorder strength ($\beta$), stress release range ($\gamma$), and system sizes ($L$). In section \ref{LLS}, we explore the local load sharing limit of the model which can be achieved by setting a very high $\gamma$. On the other hand, section \ref{gamma}, deals with a generalized version of the model where the stress of a broken fiber is redistributed depending on the exponent $\gamma$ keeping $\beta$ constant. $\beta$ is varied between 0.4 and 2.0 while $\gamma$ varies from $0$ to $3$. The size of the bundle varies from $10^3$ to $10^5$. $10^4$ realizations (bundle replications) are considered for our numerical simulation. Section \ref{universality} shows numerical results for uniform and Weibull threshold distribution, discussing the universal behavior of our result. Finally, in section \ref{discussion} we have provided the concluding remarks on the present article.


\subsection{Local Load Sharing Fiber Bundle Model: Variation in $\beta$}\label{LLS}

In this section, we have studied the fiber bundle model in the local load sharing limit. In this limit, the stress of a broken fiber is redistributed among the nearest neighbors only. This scenario can be achieved by setting a very high value of $\gamma$.

\begin{figure}[ht]
\centering
\includegraphics[width=8.0cm]{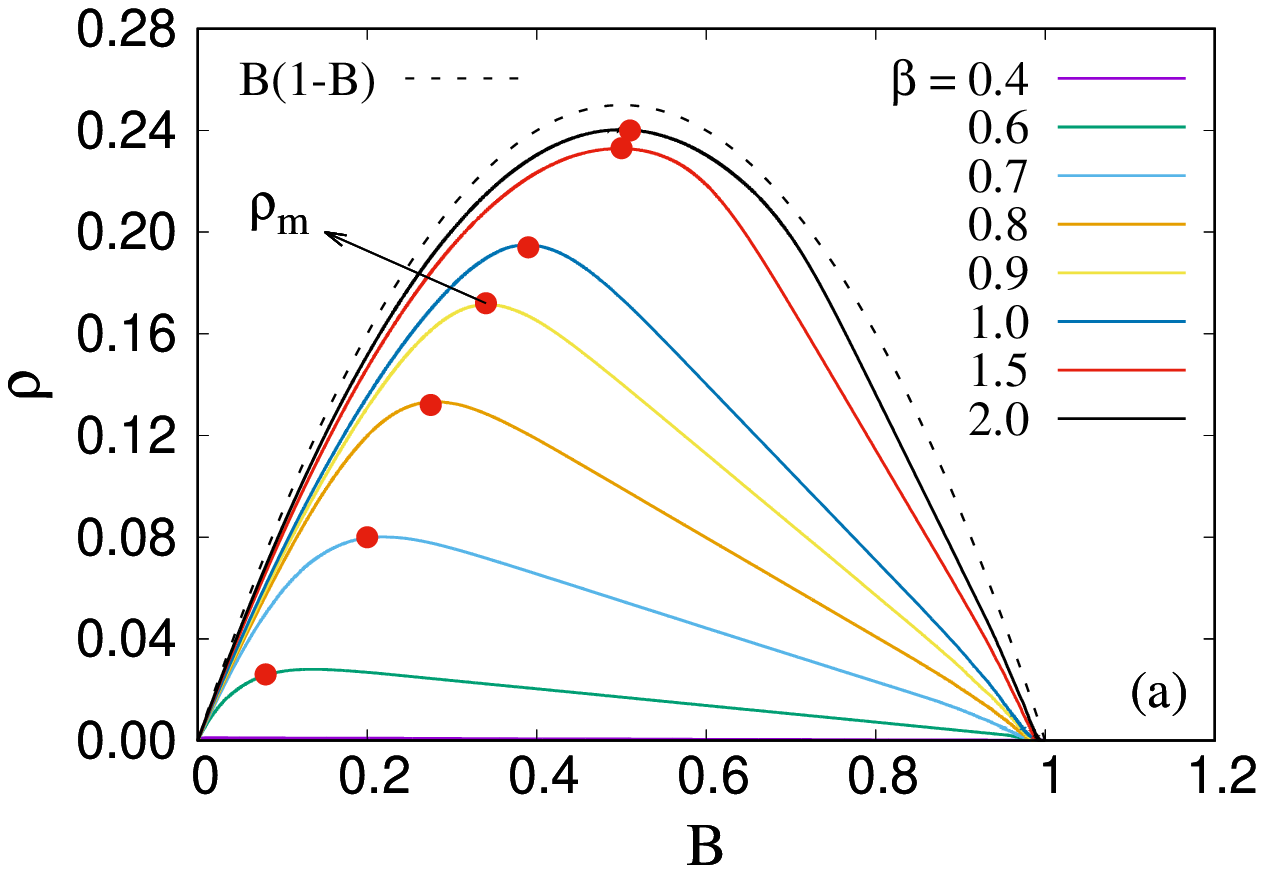} \\ \includegraphics[width=8.0cm]{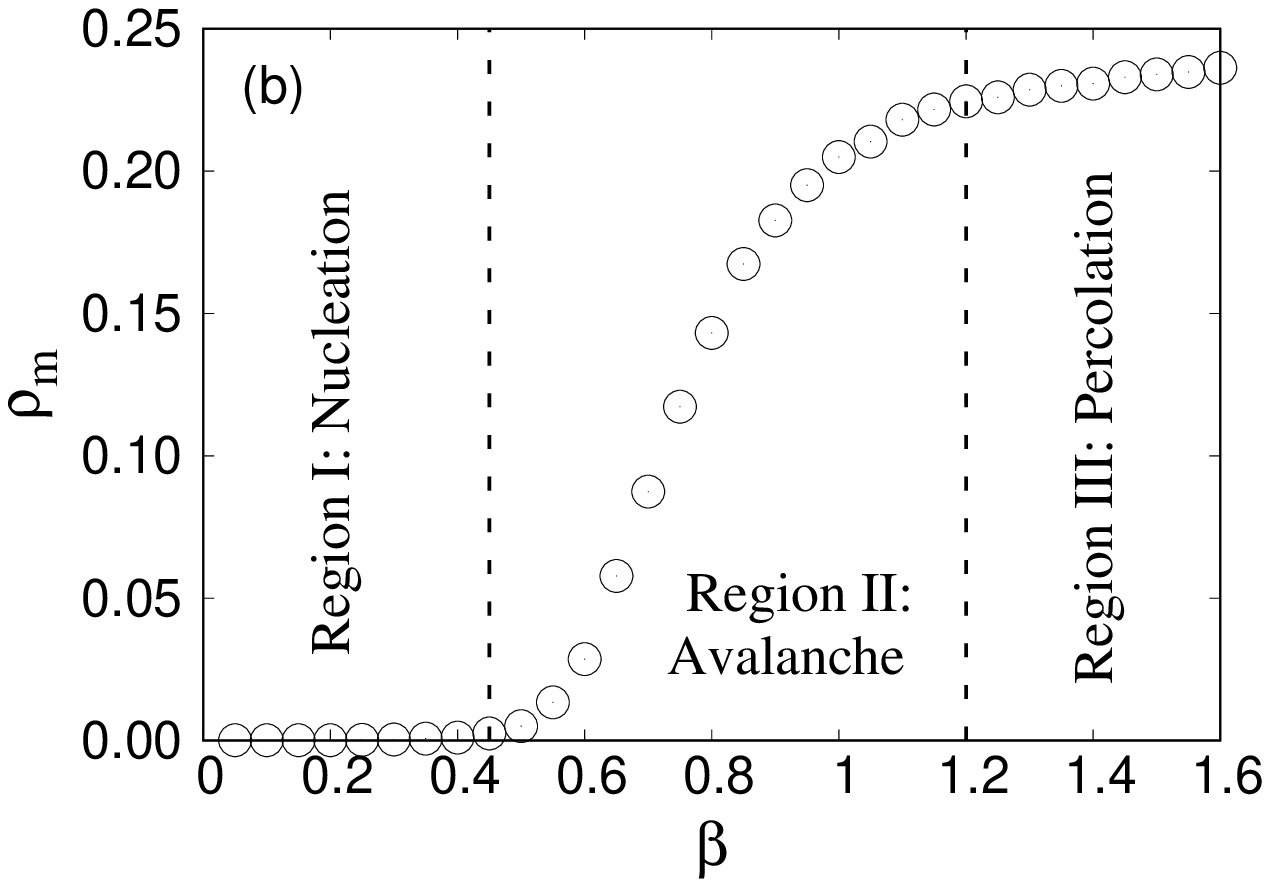}
\caption{(a) Variation of patch density $\rho$ with fraction broken $B$ for $\beta$ values ranging from 0.4 to 2.0. The red dots denotes $\rho_m$, the maximum possible patch density. The dotted line is the locus of $B(1-B)$ and represents random failure events for a 1d bundle. \\ (b) Variation of $\rho_m$ as the disorder strength $\beta$ is varied continuously. We see three different regions. (I) $\rho_m$ is zero here suggesting a single crack propagates through the bundle in a nucleating manner. (II) Here number of cracks originates and the number increases with $\beta$. (III) The failure process here random in space. $\rho_m$ has a value close to 0.25 here. The system size is kept constant at $10^5$.}
\label{fig1}
\end{figure}

We start our numerical simulation by observing the characteristics of the patches (or cracks) that are generated within the bundle during the evolution of the model. A certain patch is defined by the combination of an intact fiber and a broken fiber in its neighborhood. If we denote the intact fiber by 1 and broken fiber by 0, then any (1,0) or (0,1) combination on the one-dimensional chain of fibers will be characterized as a patch or crack. The patch density $\rho$ at time $t$ is defined by the number of patches at that time, normalized by the size $L$ of the system. We also note the fraction $B$ of broken fibers at time $t$. $B$ is defined as the number of broken fibers divided by size $L$ of the system. {\it Time}, in this case, is represented as the sum of stress increment and redistribution steps (see ref \cite{time} for details). Figure \ref{fig1} shows the variation of $\rho$ with $B$ for different disorder strength values ranging in between 0.4 and 2.0 for a bundle of size $L=10^5$. 

Figure \ref{fig1}(a) shows a non-monotonic behavior of $\rho$ when $B$ increases from 0 to 1. Earlier papers \cite{brr15,rbr17} have already discussed that such crack density shows a non-monotonic behavior as the model evolves. $B=0$ stands for the initial configuration where all fibers are intact while we obtain $B=1$ when all fibers are broken. The patch density $\rho$ is zero for $B=0$ as no cracks are there in the bundle. On the other hand, when the model is close to the failure point, there will be a single patch, making $\rho=1/L$ when $B$ approaches 1. At an intermediate $B$, $\rho$ reaches a maximum value $\rho_m$. Before this maxima, $\rho$ is an increasing function of $B$ as new patches are generated within the bundle. After the maxima, the patches start to coalesce with each other and we observe a lesser and lesser number of cracks with increasing time (hence increasing $B$). Figure \ref{fig1}(a) shows as disorder strength is increased, $\rho_m$ shifts to a higher value. We will discuss the variation of $\rho_m$ with disorder next. The dotted line in the same figure is the locus of $\rho = B(1-B)$. This dotted line represents failure events random in space for a 1d FBM. This can be understood as follows. If $B$ is the fraction of fibers broken, then the probability of having a broken and an intact fiber will be $B$ and $(1-B)$ respectively. A patch then will be created by placing an intact fiber beside a broken one, the probability of which will be $B(1-B)$ on a 1d lattice. By equating $d\rho/dB$ for this locus to zero we get the maximum $\rho_m=0.25$ and the $B$ value to be 0.5 at this maxima. The failure pattern becomes more and more random when $\beta$, the disorder strength, is high enough. On the other hand, when the disorder strength is very low ($\beta=0.4$), we see $\rho=1/L$ independent of the value of $B$. This suggests a pure nucleating failure starting from the very beginning until the global failure. 

Figure \ref{fig1}(b) shows the variation of $\rho_m$ explicitly when $\beta$ is continuously varied. We observe three different regions. 

{\bf (I) Nucleation ($\beta \le 0.4$) $-$} Here $\rho_m$ has a value close to $1/L$. This suggests that only a single crack is generated within the bundle in this limit and this crack nucleates to create global failure. Due to the low strength of the disorder, the failure process here is guided by the local stress concentration at the crack tips. 

{\bf (III) Percolation ($\beta \ge 1.2$) $-$} In this limit the behavior of $B$ vs $\rho$ matches closely with $\rho=B(1-B)$. $\rho_m$ has a value close to $0.25$. The failure events are random in space here making it reminiscent of percolation in 1d lattice. The failure process is completely guided by the disorder strength and the local stress concentration is almost non-existing. 

{\bf (II) Avalanche ($0.4 < \beta <1.2$) $-$} in the intermediate disorder strength, there is an interplay between the disorder strength and the local stress concentration. The failure process here starts in a percolating manner but later the local stress concentration takes over making the rest of the failure events nucleating. We will be discussing this spatial correlation in more detail later in this paper. \\

\begin{figure}[ht]
\centering
\includegraphics[width=8.0cm]{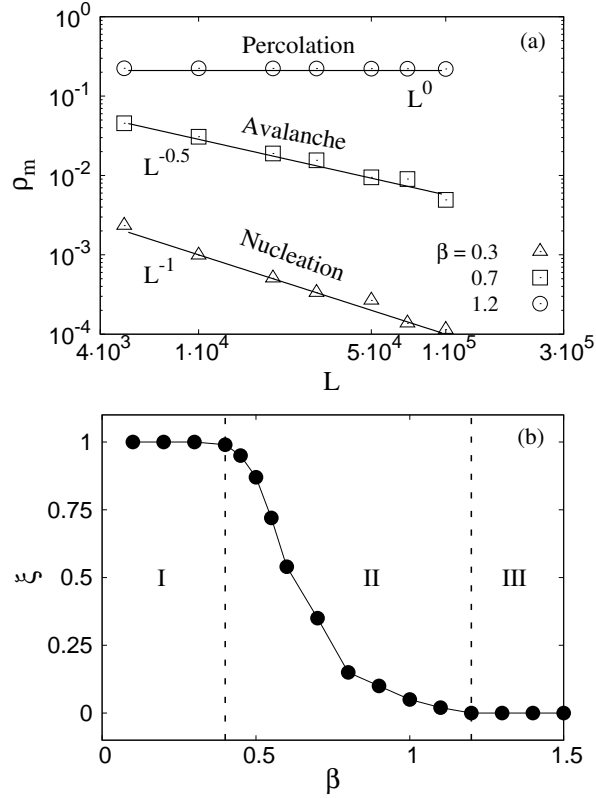}
\caption{(a) Effect of system size $L$ on $\rho_m$ for $\beta=0.3$, 0.7 and 1.2. $L$ is varied in between $5\times10^3$ and $10^5$. Three regions are observed here. (I) Nucleation: $\rho_m \sim L^{-1}$, (II) Avalanche: $\rho_m \sim L^{-\xi(\beta)}$, (III) Percolation: $\rho_m \sim L^0$. \\ (b) Variation of $\xi$ with $\beta$. $\xi=1$ and 0 in the region I and III respectively. In the intermediate region II, $\xi$ decreases continuously with $\beta$.}
\label{fig2}
\end{figure}

Figure \ref{fig2}(a) shows how the maximum patch density $\rho_m$ responds to the size of the bundle. The results are repeated for three different $\beta$ values 0.3, 0.7, and 1.2, in order to cover all three regions$-$ nucleation, avalanche, and percolation, mentioned above. When disorder strength $\beta$ is low (0.3), $\rho_m$ decreases with $L$ in a scale-free manner with exponent $-1$. This suggests that the maximum number of cracks observed in the bundle decreases with increasing size and the model goes towards nucleation more and more as the model approaches the thermodynamic limit. On the other hand, for $\beta=1.2$, $\rho_m$ is independent of $L$ and saturates at a value close to 0.25. As mentioned above, the failure process is percolating here and remains the same irrespective of the size of the bundle. In the intermediate disorder, where the disorder strength and local stress concentration compete with each other, we observe
\begin{align}
\rho_m \sim L^{-\xi(\beta)}
\end{align} 
where the exponent $\xi$ is a function of $\beta$.

Figure \ref{fig2}(b) shows the variation of exponent $\xi$ with the strength of disorder $\beta$. $\xi$ remains at 1 for low $\beta$ where the failure is nucleating, gradually decreases in region avalanche, and becomes constant at 0 in the limit of percolation. The nature of patch density remains the same in the percolation region ($\xi=0$) only. In both avalanche and nucleation, fewer patches grow as the size of the bundle is increased, suggesting that the effect of the local stress concentration becomes more prominent here as the model goes towards the thermodynamic limit. \\

\begin{figure}[ht]
\centering
\includegraphics[width=8.0cm]{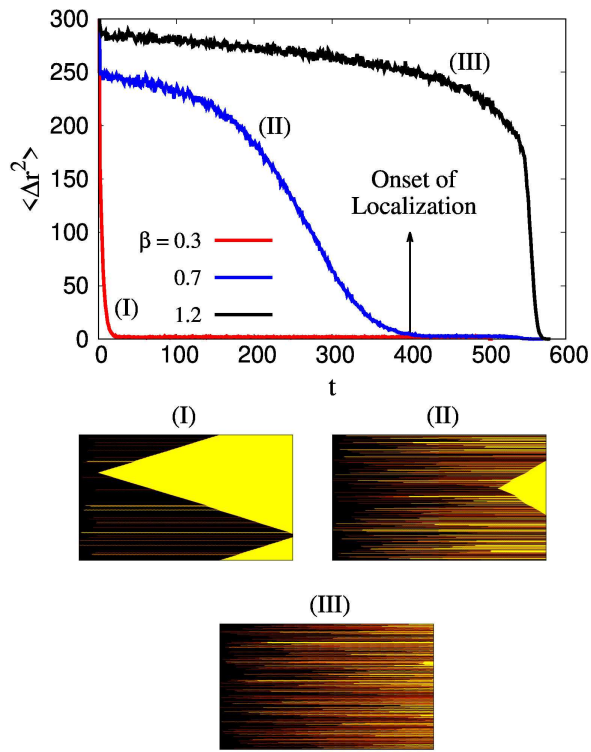}
\caption{Variation of $\langle \Delta r^2 \rangle$ with time $t$ for three different disorder strengths $\beta=0.3$, 0.7 and 1.2. $\langle \Delta r^2 \rangle$ is average square distance between consecutive rupture events. $\langle \Delta r^2 \rangle$ starts from a high value and decreases to 1 with time suggesting pure nucleation beyond this point. The spatial correlation is shown explicitly for three different cases: (I) nucleation, (II) avalanche and (III) percolation.}
\label{fig3}
\end{figure}

Here, we will discuss a dynamic parameter that helps us to understand the onset of the nucleation process with time more clearly. As explained earlier, time $t$ here is analogous to the total number of redistribution plus stress increment steps prior to the global failure. We start by breaking the weakest fiber, say $i$, at time $t=0$ by the first stress increment. Let us assume further $n_1$ fibers break at the next time step ($t=1$) upon redistributing the stress carried by the weakest fiber. We consider the distance $\Delta r$ between these two consecutive events to be the minimum of distances between fiber $i$ and other $n_1$ fibers that break after redistribution. Here, $\Delta r$ is not the exact lattice distance as only intact fibers are considered while calculating it. The distance across a broken patch is considered to be 1 independent of the size of the patch. This is due to the LLS scheme that we have adopted. Whenever a fiber at a notch breaks and the redistributed stress breaks the fiber at the other notch, the failure is still nucleating, no matter how large this patch is. Next, we square this distance and average it over $10^4$ realizations to get $\langle \Delta r^2 \rangle$ at time $t=0$. Next, we move our reference frame to the fiber among those $n_1$ fibers that had the minimum distance from fiber $i$. Let's denote this new fiber as $j$. If further $n_2$ fibers break in the next redistribution, $\langle \Delta r^2 \rangle$ at $t=1$ will be calculated by the same procedure: find $\Delta r$ from the minimum of distances between fiber $j$ and those $n_2$ fibers, square it and average over $10^4$ realizations. Such a parameter was explored earlier by Stormo et. al \cite{sgh12} in the context of the soft clamp model to point out the onset of localization. Figure \ref{fig3} shows this variation of $\langle \Delta r^2 \rangle$ with time $t$ for $\beta=0.3$, 0.7 and 1.2. For all three disorder strength values, $\langle \Delta r^2 \rangle$ starts from a high value at low $t$ and then decreases towards 1 when $t$ is high. A high value of $\langle \Delta r^2 \rangle$ suggests the fibers that break consecutively are far from each other. This is a spatially uncorrelated failure. On the other hand, when $\langle \Delta r^2 \rangle=1$, the consecutive failures take place from the neighboring fibers only. This behavior stands for pure nucleation. For $\beta=0.3$, $\langle \Delta r^2 \rangle$ becomes 1 very fast and stays there independent of $t$ until the bundle reaches global failure. For $\beta=1.2$, we observe the opposite behavior where $\langle \Delta r^2 \rangle$ stays at a high value for a long time and falls to 1 just before global failure. The former behavior is nucleating (I) while the latter one is percolating (III). The visualization of the failure process for both (I) and (III) is shown below figure \ref{fig3}. The x-axis of each plot is time and the y-axis is fiber index. The color gradient is over the local stress profile. The yellow color stands for the failed fibers. For (I), we see a single crack growing in a nucleating manner from the very beginning. For (III), on the other hand, there are no nucleating yellow-colored fibers and the rupture events are spatially uncorrelated. For the avalanche (II) behavior, there is a spatially uncorrelated failure in the beginning as well as nucleation close to global failure. Figure \ref{fig3} shows the point for $\beta=0.7$ where $\langle \Delta r^2 \rangle$ becomes 1 indicating onset of localized (nucleating) failure events.  \\

\begin{figure}[ht]
\centering
\includegraphics[width=8.5cm]{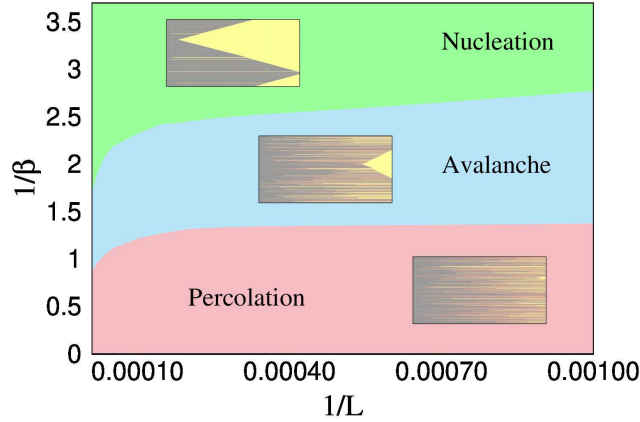}
\caption{Phase diagram of disorder strength $\beta$ and system size $L$. $1/\beta$ is plotted against $1/L$ to make the coordinate (0,0) represent infinite disorder at thermodynamic limit. Three regions $-$ nucleation, avalanche and percolation are observed. The spatial correlation during fiber rupture is shown with yellow being the broken patch.}
\label{fig4}
\end{figure}

Finally, we have constructed the phase diagram of disorder strength $\beta$ and system size $L$ to show all three failure processes. In figure \ref{fig4}, $1/\beta$ is plotted against $1/L$. This is done in this way so that the origin (0,0) of this plot corresponds to $L\rightarrow\infty$ and $\beta\rightarrow\infty$, an infinite disorder in the thermodynamic limit. As discussed earlier, if the disorder strength is increased, we start with nucleation, go through an avalanche, and finally reach percolation behavior. The spatial rupturing events are shown in the corresponding phases. Now, if the disorder strength is kept constant and the size of the bundle is increased, a percolating behavior moves towards avalanche and an avalanche behavior moves towards nucleation. Due to weak dependence of parameters like $\rho_m$ on $L$ (see figure \ref{fig2}), it will not be possible to see (as the system size has to be very high) this change from percolation to avalanche if we are well inside the percolation region. To see this change at relatively lower system sizes, it is required to keep the disorder strength at a value so that the model is closer to the percolation-avalanche interface. The opposite happens if the system size is decreased instead of increasing. This suggests we observe only nucleating failure in the thermodynamic limit unless the disorder is infinitely high. This effect of disorder was explored earlier in the context of random fuse network by Shekhawat et al. \cite{szs13} and Moreira et al. \cite{mohaha12}. We observe that the fiber bundle model in one dimension follows the same trend. \\


\subsection{Generalized Model: Variation in both $\beta$ and $\gamma$}\label{gamma}

In this section, the model is studied with a continuous variation in both $\gamma$ when $\beta$. We start our numerical simulation by observing the spatial correlation through the rupture events with increasing time as the bundle fails. \\

\begin{figure}[ht]
\centering
\includegraphics[width=11cm]{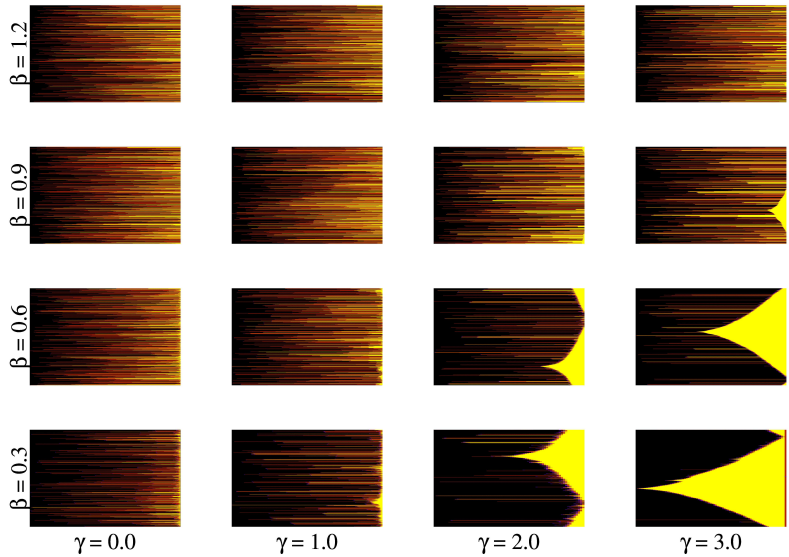} 
\caption{The figure shows spatial correlation during a failure process when $\beta$ varies between 0.3 and 1.2 and $\gamma$ varies between 0 and 3. Nucleating failure is observed for high $\gamma$ and low $\beta$. The spatial correlation reduces as $\gamma$ is decreases (pushes the model towards MF limit) or $\beta$ is increased.}
\label{fig5}
\end{figure}

Figure \ref{fig5} shows such correlation for $\beta$ values ranging in between 0.3 and 1.2 and $\gamma$ values within $0$ and $3$. For each small figure, the x-axis shows the time $t$ and the y-axis shows the fiber index. From left to right, the figures are plotted for increasing values of $\gamma$ keeping the disorder strength $\beta$ constant. On the other hand, from bottom to the top the figures are plotted with increasing $\beta$ and keeping $\gamma$ constant. We observe the following behavior: 

For {\it low $\beta$} and {\it high $\gamma$}, the fibers break in a nucleating manner. Due to the low stress release range the stress concentration plays a crucial role and dominates the failure process. Moreover, due to the low value of disorder strength, the probability that the fibers break with redistribution (without any increment in external stress) from the neighborhood is high. On the other hand, for {\it high $\beta$} and {\it low $\gamma$}, the fluctuation between threshold strength as well as the stress release range is high. As a result, we observe rupture events random in space and through increment in external stress.       

Now, keeping the $\beta$ fixed at a low value, as we decrease $\gamma$, the model slowly goes towards the mean-field limit. In this limit, the stress of the broken fibers are redistributed among all surviving fibers. This increases the chance that whenever one fiber breaks, the next rupture event may take place from somewhere which is not the neighborhood of the broken fiber. The failure process, in this case gradually deviates from the nucleating behavior as $\gamma$ decreases.   

Instead, if we keep $\gamma$ fixed at a high value and increase $\beta$, the fluctuation among fiber strengths increases. Here, the stress of the broken fiber is redistributed in the neighborhood (as $\gamma$ is high) but due to this increase in fluctuation we will find more more strong fibers in this neighborhood that will finally arrest the growth of a crack - a phenomena known as lattice trapping or intrinsic crack resistance \cite{thr71,cm90,rice92,pg00,bh03,mcc05}. This forces the growth of a different crack from a different place. 

\begin{figure}[ht]
\centering
\includegraphics[width=8.0cm]{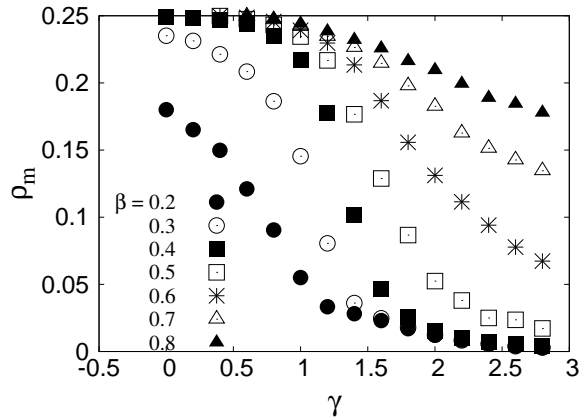}
\caption{Variation of maximum number of cracks ($\rho_m$) with $\gamma$ for $\beta$ values ranging in between 0.2 and 0.8. $\rho_m$ has a high value for low $\gamma$ and gradually decreases with $\gamma$ and tends to saturate towards a low value. For low $\beta$, $\rho_m$ decreases close to $1/L$ for high $\gamma$, suggesting propagation of a single crack. On the other hand, for low $\gamma$ and high $\beta$, $\rho_m \approx 0.25$, suggesting a failure process which is random in space.}
\label{fig6}
\end{figure}

As discussed in figure \ref{fig5}, a single crack or a number of cracks are observed in the bundle depending on what the values of disorder strength $\beta$ and the stress release range $\gamma$ are. In figure \ref{fig6}, we have studied how the maximum number of cracks ($\rho_m$) varies as we vary both $\beta$ and $\gamma$. At first, we observe the variation of $\rho_m$ with $\gamma$ for a constant value of $\beta$. The study is then repeated for $\beta$ values ranging in between $0.2$ and $0.8$. 

For low $\gamma$, $\rho_m$ has a higher value and decreases as $\gamma$ increases and crosses the critical value $\gamma_c$ \cite{brr15}. The results can be discussed in three parts. At an intermediate disorder ($0.3 \le \beta \le 0.5$), $\rho_m$ saturates close to 0.25 for low $\gamma$ and decreases to $1/L$ when $\gamma$ is high. A $\rho_m$ value close to $1/L$ suggests there is only one crack that propagates throughout the system. On the other hand, as already discussed in the manuscript, $\rho_m$ close to 0.25 suggests a failure process random in space. In this limit, all three regions are accessible with a variation in $\gamma$. 

\begin{figure}[ht]
\centering
\includegraphics[width=8.5cm]{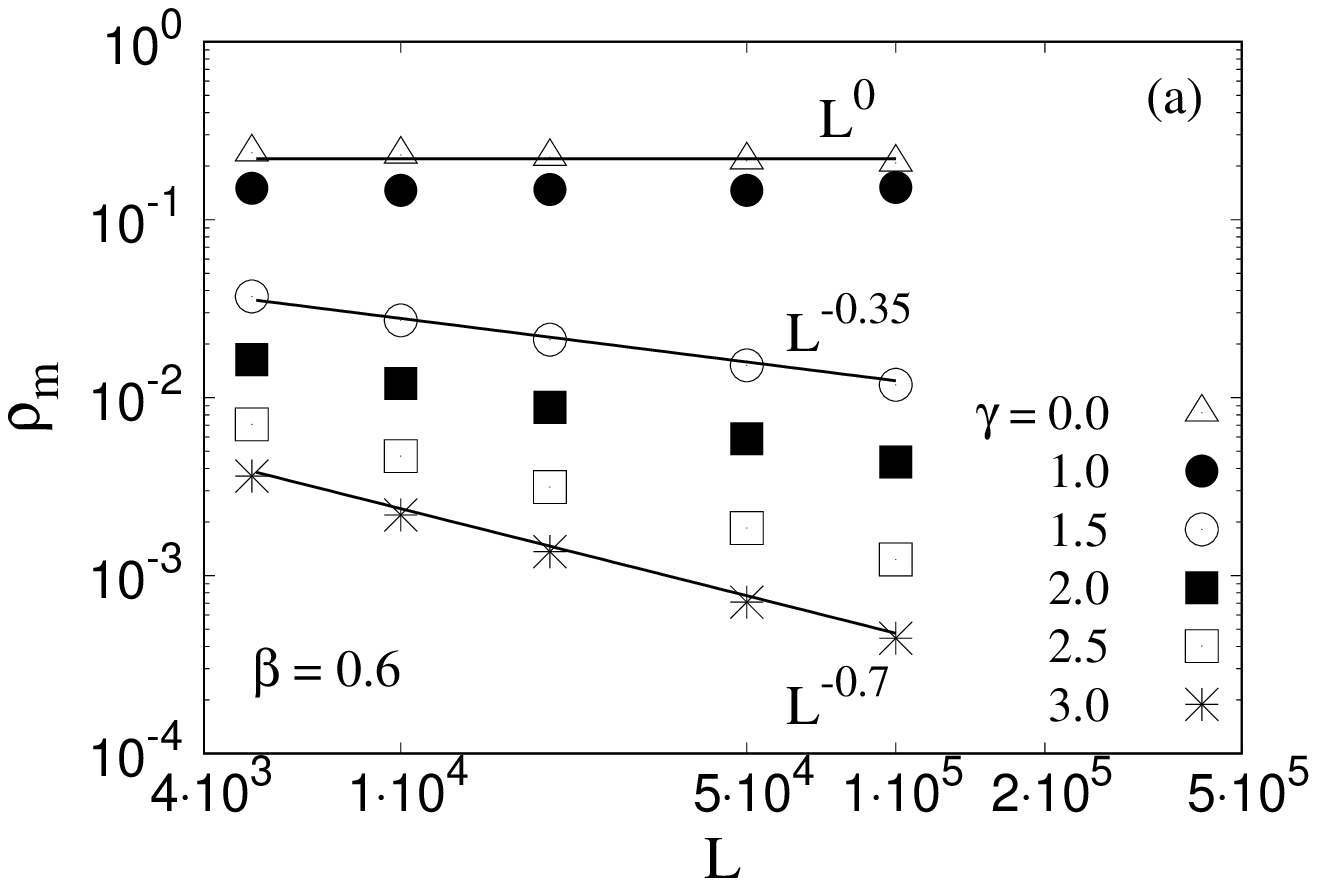} \\ \includegraphics[width=8.0cm]{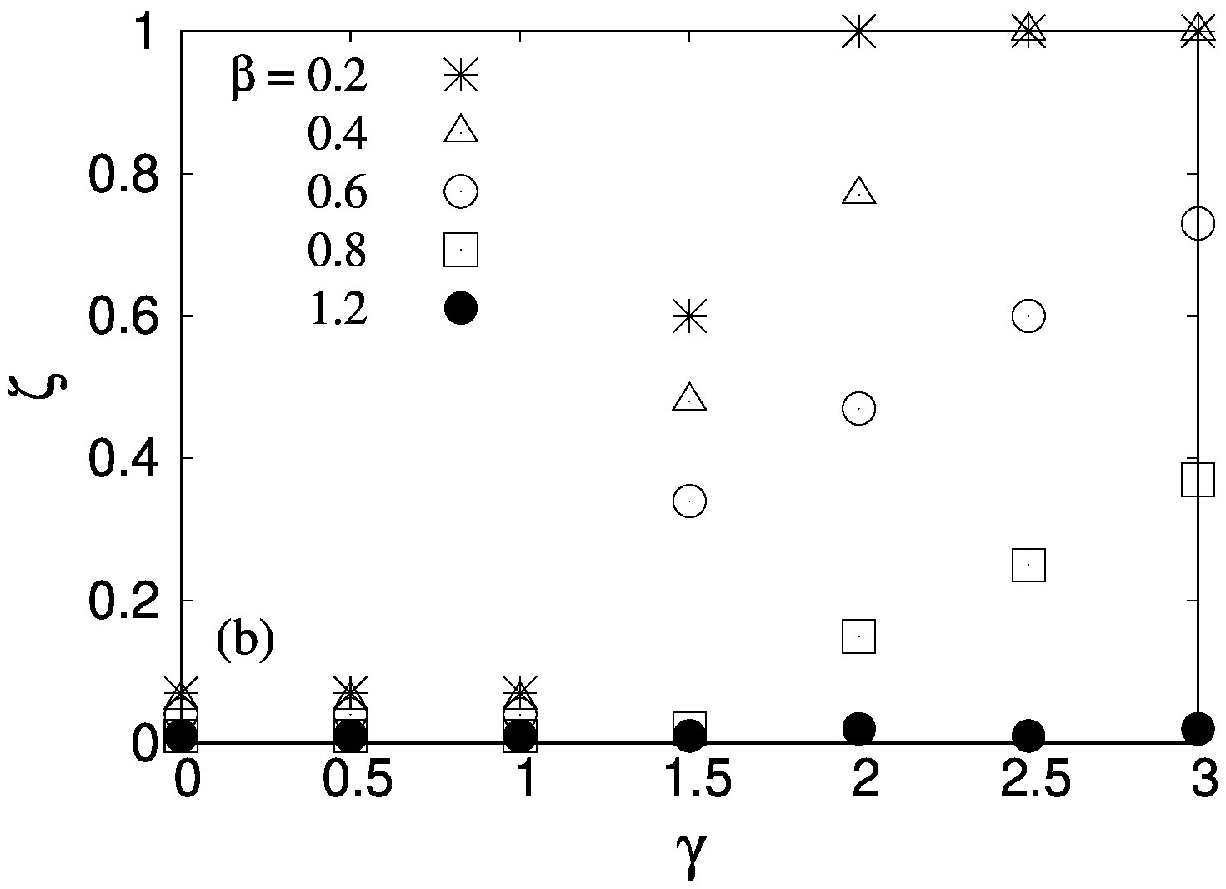}
\caption{(a) System size effect of $\rho_m$ for $\beta=0.6$ and $\gamma$ between 0 and 3. We observe $\rho_m \sim L^{-\zeta}$, where $\zeta$ is an increasing function of $\gamma$. \\ (b) Variation of $\zeta$ with $\gamma$ for $0.2 \le \beta \le 1.2$. For low $\gamma$, $\zeta \approx 0$ independent of $\beta$. For high $\gamma$, on the other hand, $\zeta=1$ depending on whether $\beta$ is low enough or not.}
\label{fig7}
\end{figure}

At low $\beta$ ($<0.3$), we observe that $\rho_m$ goes to $1/L$ for high $\gamma$ but does not approach $0.25$ even if $\gamma$ is very low. In this limit, we do not see a percolation like a random failure. This is due to very low disorder strength, where the bundle breaks very abruptly before it can reach the real maximum value of $\rho_m$ ($\approx 0.25$) at low $\gamma$. On the other hand, for high $\beta$ ($>0.5$), $\rho_m$ reaches 0.25 easily at low $\gamma$ but do not reach $1/L$ for high $\gamma$. In this case, $\rho_m$ does not reach $1/L$ even at high $\gamma$ due to the intrinsic crack resistance caused by the high fluctuation in threshold strength, which arrests an propagating crack in the process. As a result, a nucleating failure is not observed here. \\  

Figure \ref{fig7}(a) shows the system size effect of $\rho_m$ for $\beta=0.6$ and for wide range of stress relaxation $\gamma$ ($0 \le \gamma \le 3$). We observe $\rho_m$ to decrease in a scale-free manner with system size $L$,
\begin{align}
\rho_m \sim L^{-\zeta}
\end{align}
where $\zeta$ is an increasing function of $\gamma$. At low $\gamma$, $\rho_m$ is almost independent of $L$ and saturates around 0.25. When $\gamma$ is high, $\rho_m$ responds to the change in $L$ very sharply and decreases as $L$ increases. 

The variation of the exponent $\zeta$ is shown in figure \ref{fig7}(b). $\zeta$ has a value close to $0$ independent of disorder strength $\beta$ when $\gamma$ low. At such a low value of $\gamma$, the model is in the mean field limit and changing the system size does not change the dynamics of the model. As $\gamma$ increases, the model slowly deviates from the mean-field limit and the effect of local stress concentration becomes more and more prominent. In this limit, $\zeta$ starts to increase slowly. When $\gamma$ crosses a certain value that depends on $\beta$, $\zeta$ finally reaches 1. A Higher value of $\beta$ will require a higher $\gamma$ value in order to obtain $\zeta=1$. Finally, when $\beta$ is very high, $\zeta$ remains close to $0$ independent of the stress release range $\gamma$. Here, the failure process is random in space, independent of both $\beta$ and $\gamma$. \\

\begin{figure}[ht]
\centering
\includegraphics[width=8.5cm]{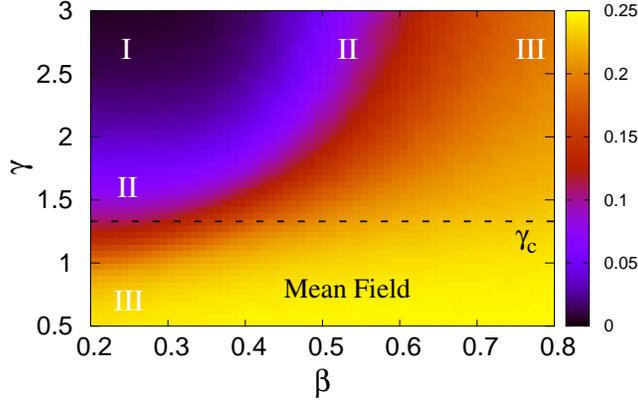} 
\caption{Histogram of $\rho_m$ on the plane of $\gamma$ vs $\beta$. The brightest color corresponds to $\rho_m \approx 0.25$ while the darkest color stands for $\rho_m \approx 1/L$. At low $\beta$, as $\gamma$ increases, we go from region III (percolation) to II (avalanche) to I (nucleation). Same transition is observed when $\beta$ decreases at high $\gamma$.}
\label{fig9}
\end{figure}

Figure \ref{fig9} shows above mentioned three regions from the study of the maximum number of cracks ($\rho_m$) when both disorder strength $\beta$ and the stress release range $\gamma$ are varied simultaneously. The color gradient in figure \ref{fig9} is on $\rho_m$, with a maximum value of 0.25 (lightest color) and minimum value of $1/L$ (darkest color) which is $10^{-5}$ as the size of our bundle is $10^5$.   
We observe $\rho_m$ to have the lowest value for lowest possible $\beta$ and highest possible $\gamma$. 

This is due to the fact that fluctuation among local strength values are minimum here and at the same time stress release range is also minimum making the local stress concentration most prominent. At this situation, as we decrease $\gamma$, we go towards percolation (light color) through avalanche region. The same behavior is observed (I $\rightarrow$ II $\rightarrow$ III) if we increase $\beta$ keeping $\gamma$ fix. The figure also shows the existence of $\gamma_c$ \cite{brr15,hmkh02}, the value of $\gamma$ below which the model enters the mean-field limit. For $\gamma<\gamma_c$, we have almost an uniform gradient of light color suggesting $\rho_m$ is close to $0.25$ here independent of $\beta$. We observe a very small change in the color gradient if we increase $\gamma$ at a high $\beta$. \\

\begin{figure}[ht]
\centering
\includegraphics[width=8.5cm]{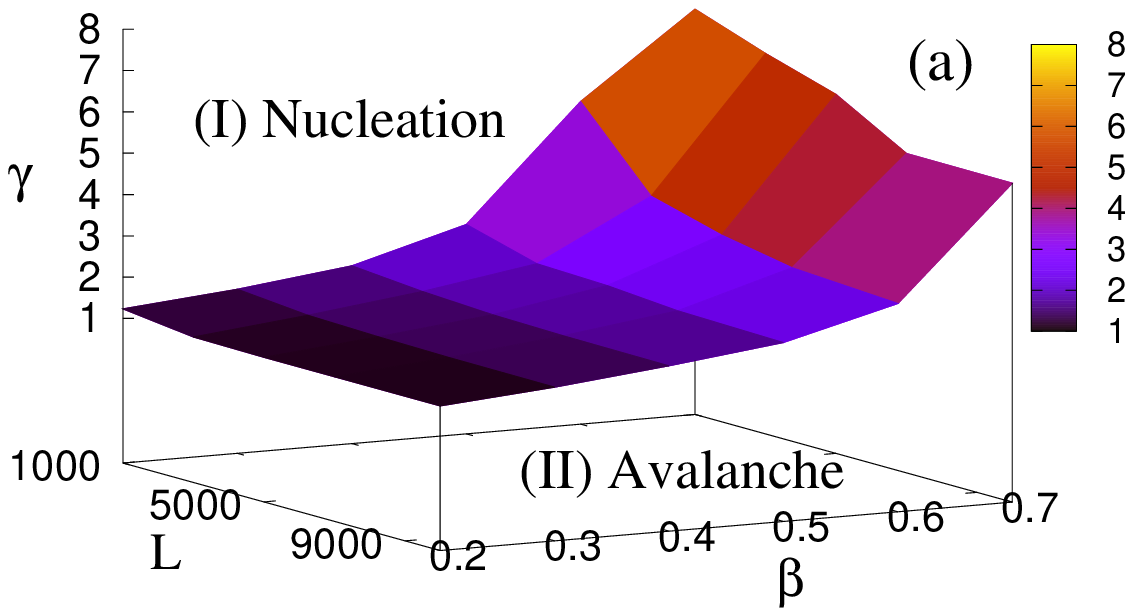} \\ \includegraphics[width=8.5cm]{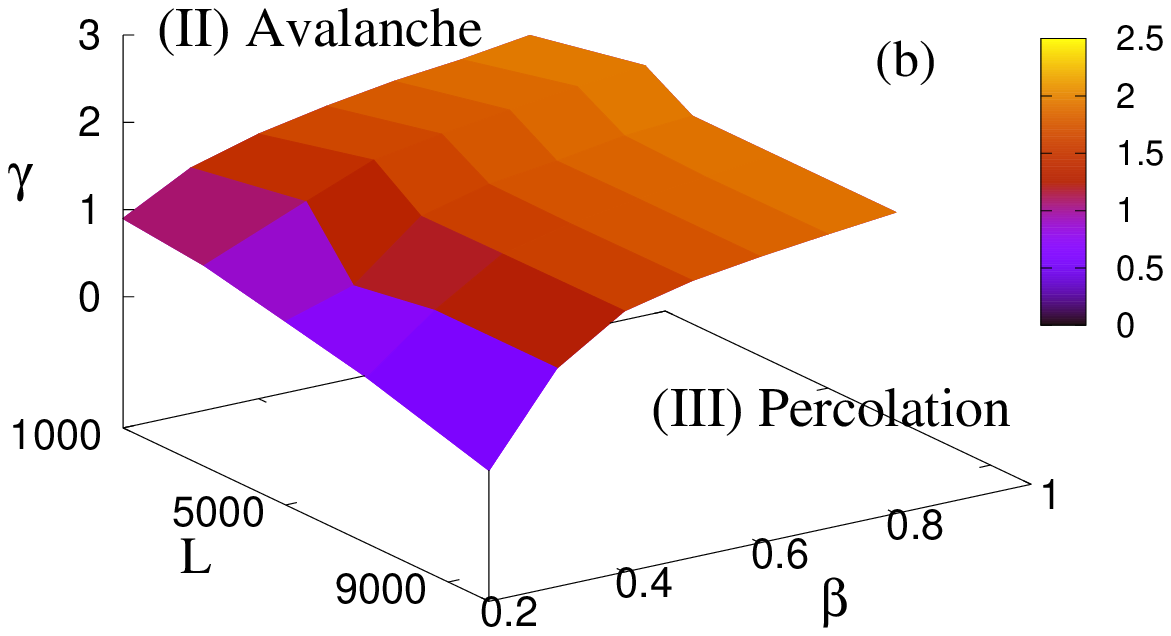} 
\caption{(a) and (b) respectively shows the plane of transition from avalanche to nucleation and percolation to avalanche respectively. The three parameters constructing the plane are: disorder strength ($\beta$), stress release range ($\gamma$) and system size ($L$).}
\label{fig10}
\end{figure}
 
Figure \ref{fig10} shows different regions $-$ nucleation, avalanche and percolation, with their unique nature of crack propagation during the failure process when the disorder strength ($\beta$), stress release range ($\gamma$), and system size ($L$) are continuously varied. Figure \ref{fig10}(a) shows the plane separating the region nucleation from avalanche. The plane seems to diverge for high $\beta$. This is due to the fact that, at high beta the fluctuation among fiber strength values will be high and the $\gamma$ value will also have to be very high to make the local stress concentration prominent enough to create nucleation. At the same time, since an increasing $L$ has already been seen to favor nucleating failure, we achieve such nucleation at relatively lower $\gamma$ value at higher $L$ when $\beta$ is kept fixed. Figure \ref{fig10}(b), on the other hand, shows the plane between avalanche and percolation. We observe the same effect of $L$ here $-$ as $L$ increases, the transition from percolation to avalanche takes place at a lower value of $\gamma$. At the same time, as $\beta$ increases, we have to go to a higher $\gamma$ value to enter the avalanche region from percolation. The sudden upward curvature of {\it nucleation $-$ avalanche} plane at high $\beta$ suggests that if the disorder strength is extremely high, we might not get a nucleation region. Similarly, the sudden downward curvature of {\it avalanche $-$ percolation} plane at low $\beta$ suggests that if the disorder strength is extremely low, we might not get a percolation region.    

\begin{figure}[ht]
\centering
\includegraphics[width=8cm]{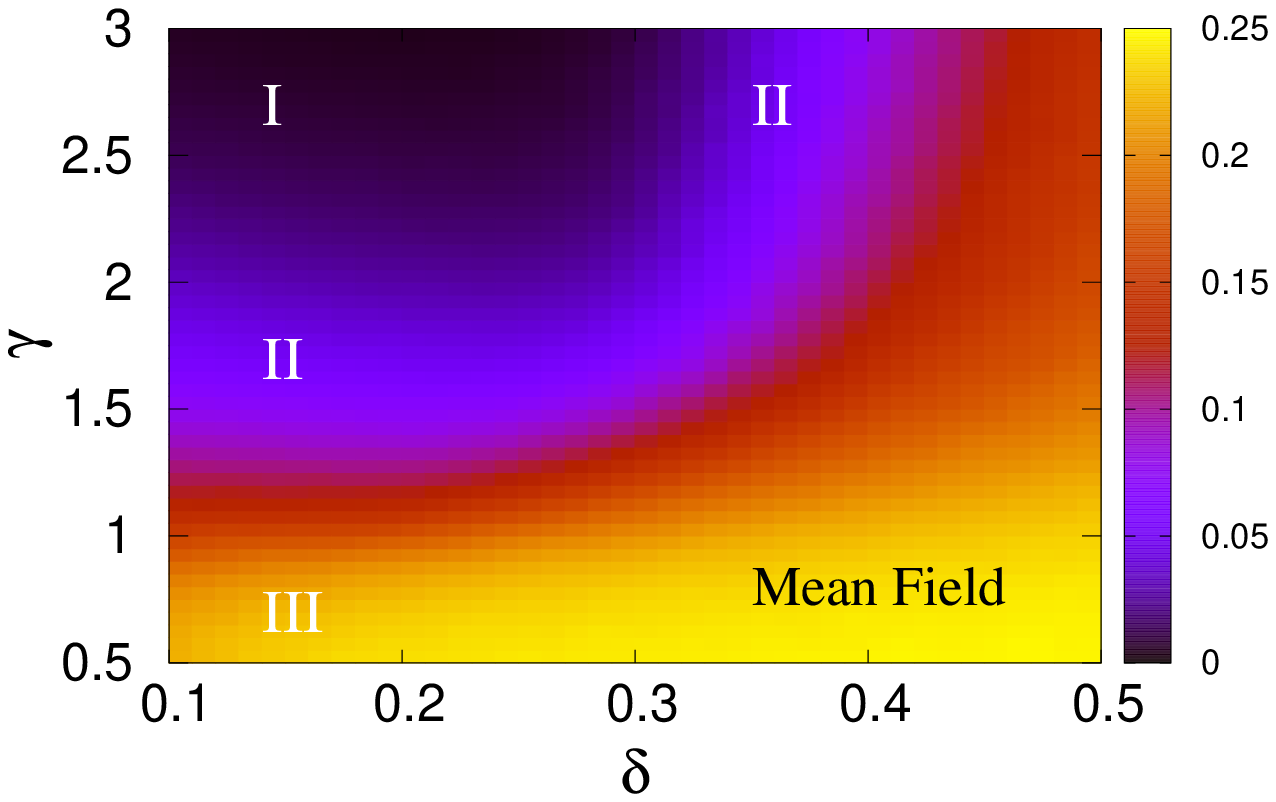} \\ \includegraphics[width=8cm]{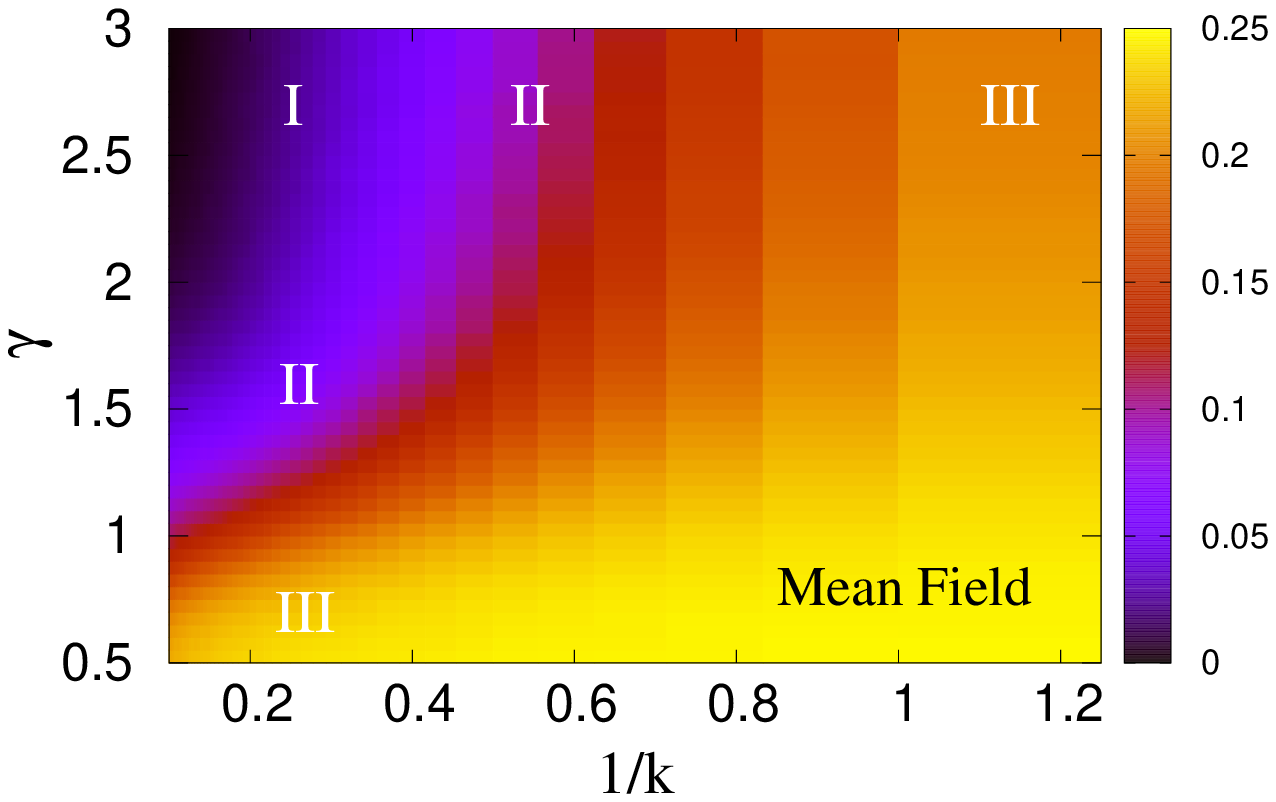} 
\caption{Variation of $\rho_m$ for uniform (upper) and Weibull (lower) distribution when both the disorder strength ($\delta$ or $k$) and stress release range ($\gamma$) is varied continuously. Three distinct regions (I) nucleation, (II) avalanche and (III) percolation, is observed independent of the choice of threshold distribution.}
\label{fig10}
\end{figure}

\section{Universality}\label{universality} 

In this section we will discuss the universality of our study by using two different distributions other than the power law. For this purpose, we have adopted an uniform and a Weibull distribution as described below:
\begin{equation}\label{equ1}
p(h) \sim \begin{cases}
    \displaystyle\frac{1}{2\delta},  & (0.5-\delta \le h \le 0.5+\delta) \\
    0.  & ({\rm otherwise})
  \end{cases}
\end{equation}
where $\delta$ is the half-width of the distribution as well as the measure of the disorder strength.
\begin{equation}\label{equ2}
p(h) \sim \left(\displaystyle\frac{k}{\lambda}\right)\left(\displaystyle\frac{h}{\lambda}\right)^{k-1}e^{-\left(\displaystyle\frac{h}{\lambda}\right)^k}
\end{equation}
where $k$ and $\lambda$ are the Weibull modulus and scale parameter respectively. $k$ controls the disorder strength in this case. We vary $\delta$ from 0 to 0.5 while $k$ is varied between $1$ and $10$. The scale parameter $\lambda$ is kept constant at $1$. Similar to the power law distribution, the uniform distribution is also a bounded distribution while the Weibull distribution is open.

Figure\ref{fig10} shows the variation of $\rho_m$ when both disorder strength and stress release range is tuned continuously. The results for uniform distribution are shown in the left figure. The right figure shows the same for the Weibull distribution. Similar to figure \ref{fig9}, the color gradient represents the variation in $\rho_m$ that spans from $1/L$ to $0.25$. The results suggest that all three regions: (I) nucleation, (II) avalanche, and (III) percolation, are observed independent of the choice of the threshold distribution. As $\gamma$ increases, the model goes from the nucleating failure to a failure process random in space. On the other hand, a spatially correlated failure process is not observed by increasing $\delta$ up to 0.5 (the distribution spans from 0 to 1) as the disorder strength is not large enough. Though for Weibull distribution, both nucleating and random failure is observed at high and low $k$ respectively.     


\section{Discussions}\label{discussion}

As we have already discussed, two major factors governing the mode of failure in disordered solids are the strength of heterogeneities and the effective range over which the stress field is modified following a local rupture event. On the other hand, studies in random resistor network model \cite{szs13,mohaha12} claims the failure mode, in the large system size limit, to be always nucleation-driven unless the strength of disorder is extremely high. Qualitatively this is the main finding of the present paper as well as what was observed in the random fuse network model earlier \cite{szs13,mohaha12}. The simplicity of the fiber bundle model allows us to include extra parameters like stress release range compared to the random fuse network model and study its effect as well on the spatial correlation as the model evolves. The precursor events (such as scale-free size distribution of rupture events prior to global failure and scale free distribution of emitted energies during such avalanches), previously seen in the statistical models \cite{anz06,rb21,bbr21}, would imply that a nucleation-like failure would not be achievable even in the large system size limit. Such precursor events are observed experimentally \cite{bcipsssv13} as well for which the extreme disorder is not necessarily the physical condition. The stress release range (analogous to fracture process zone in real experiments) comes into play here that might cause a different mode of failure, other than nucleation, even when the system size is high. 

In conclusion, we present a detailed study in fiber bundle model by varying main three parameters, strength of disorder, range of stress relaxation and system size, that determines the dynamics of the model as it is acted by an external stress. An increasing disorder strength (increasing $\beta$) or stress release range (decreasing $\gamma$) favors a failure that is random in space. On the other hand an increasing system size makes the failure more and more nucleating. The avalanche behavior is observed for all $\beta$ or $\gamma$. If $\beta$ is very high then it is difficult to achieve the nucleating behavior unless the value of $\gamma$ is very high. On the other hand, when $\beta$ is low, achieving nucleating failure is easy but it is difficult to observe pure random failure by decreasing $\gamma$. Finally, for the intermediate $\beta$ value, we achieve both nucleation and percolation like failure by increasing and decreasing $\gamma$ respectively.  


\section{Acknowledgment}

The work was supported by the Research Council of Norway through its Centres of Excellence funding scheme, project number 262644. 



\begin{thebibliography}{99}
\bibitem{citation1} \textit{Mechanical Metallurgy}, George E. Dieter, Metric Editions, Materials Science \& Metallurgy.
\bibitem{citation2} \textit{Fracture of Brittle Solids}, Brian Lawn, Cambridge Solid State Science Series.
\bibitem{knott73} J. F. Knott, {\it Fundamentals of Fracture Mechanics}, Butterworths, 1973.
\bibitem{Phoenix} S. L. Phoenix, \href{https://www.cambridge.org/core/journals/advances-in-applied-probability/article/abs/asymptotic-distribution-for-the-time-to-failure-of-a-fiber-bundle/DA249157868C46F75D98E51F12FC6778}{Adv. Appl. Probab. {\bf 11}, 153 (1979)}.
\bibitem{hp78} D. G. Harlow, and S. L. Phoenix, \href{https://journals.sagepub.com/doi/10.1177/002199837801200207}{Journal of Comp. Mat. {\bf 12}, 195 (1978)}.
\bibitem{bazabt04} Z. P. Bazant, \href{https://www.pnas.org/content/101/37/13400}{PNAS {\bf 101}, 13400 (2004)}.
\bibitem{anz06} M. J. Alava, P. K. V. V. Nukala, and S. Zapperi, \href{https://www.tandfonline.com/doi/abs/10.1080/00018730300741518}{Advances in Physics {\bf 55}, 349 (2006)}.
\bibitem{hh92} P. C. Hemmer, and A. Hansen, \href{https://asmedigitalcollection.asme.org/appliedmechanics/article-abstract/59/4/909/389551/The-Distribution-of-Simultaneous-Fiber-Failures-in?redirectedFrom=fulltext}{Journal of applied mechanics {\bf 59}, 909 (1992)}.
\bibitem{ppvac94} A. Petri, G. Paparo, A. Vespignani, A. Alippi, and M. Costantini, \href{https://journals.aps.org/prl/abstract/10.1103/PhysRevLett.73.3423}{Phys. Rev. Lett. {\bf 73}, 3423 (1994)}.
\bibitem{ggbc97} A. Garcimartin, A. Guarino, L. Bellon, and S. Ciliberto, \href{https://journals.aps.org/prl/abstract/10.1103/PhysRevLett.79.3202}{Phys. Rev. Lett. {\bf 79}, 3202 (1997)}.
\bibitem{sta02} L. I. Salminen, A. I. Tolvanen, and M. J. Alava, \href{https://journals.aps.org/prl/abstract/10.1103/PhysRevLett.89.185503}{Phys.Rev. Lett. {\bf 89}, 185503 (2002)}.
\bibitem{bb11} D. Bonamy, and E. Bouchaud, \href{https://www.sciencedirect.com/science/article/abs/pii/S0370157310002115}{Physics Reports {\bf 498}, 1 (2011)}.
\bibitem{dbl86} P. M. Duxbury, P. D. Beale, and P. L. Leath, \href{https://journals.aps.org/prl/abstract/10.1103/PhysRevLett.57.1052}{Phys. Rev.Lett. {\bf 57}, 1052 (1986)}.
\bibitem{dbl87} P. M. Duxbury, P. L. Leath, and P. D. Beale, \href{https://pubmed.ncbi.nlm.nih.gov/9942053/}{Phys. Rev.B {\bf 36}, 367 (1987)}.
\bibitem{cb97} B. K. Chakrabarti, and L. G. Benguigui, {\it Statistical Physics of Fracture and Breakdown in Disordered Systems}, Oxford Science Publications, Oxford, (1997).
\bibitem{msnasz12} C. Manzato, A. Shekhawat, P. K. V. V. Nukala, M. J.Alava, J. P. Sethna, and S. Zapperi, \href{https://journals.aps.org/prl/abstract/10.1103/PhysRevLett.108.065504}{Phys. Rev. Lett. {\bf 108}, 065504 (2012)}.
\bibitem{rahg88} S. Roux, A. Hansen, H. Herrmann, and E. Guyon, \href{https://link.springer.com/article/10.1007/BF01016411}{Journal of Statistical Physics {\bf 52}, 237 (1988)}.
\bibitem{hs03} A. Hansen, and J. Schmittbuhl, \href{https://journals.aps.org/prl/abstract/10.1103/PhysRevLett.90.045504}{Phys. Rev. Lett. {\bf 90}, 045504 (2003)}.
\bibitem{thr71} R. Thomson, C. Hsieh and V. Rana, \href{https://aip.scitation.org/doi/10.1063/1.1660699}{J. Appl. Phys. {\bf 42} , 3154 (1971)}.
\bibitem{cm90} W. A. Curtin, \href{https://link.springer.com/article/10.1557/JMR.1990.1549}{J. Mater. Res. {\bf 5}, 1549 (1990)}.
\bibitem{rice92} J. R. Rice, \href{https://www.sciencedirect.com/science/article/abs/pii/S0022509605800122}{J. mech. Phys. Solids {\bf 40}, 239 (1992)}.
\bibitem{pg00} R. Perez, and P. Gumbsch, Phys. \href{https://journals.aps.org/prl/abstract/10.1103/PhysRevLett.84.5347}{Rev. Lett. {\bf 84}, 5347 (2000)}.
\bibitem{bh03} N. Bernstein, and D. W. Hess, \href{https://journals.aps.org/prl/abstract/10.1103/PhysRevLett.91.025501}{Phys. Rev. Lett. {\bf 91}, 025501 (2003)}.
\bibitem{mcc05} A. Mattoni, L. Colombo, and F. Cleri, \href{https://journals.aps.org/prl/abstract/10.1103/PhysRevLett.95.115501}{Phys. Rev. Lett. {\bf 95}, 115501 (2005)}.
\bibitem{hh94} A. Hansen, and P. C. Hemmer, \href{https://www.sciencedirect.com/science/article/abs/pii/0375960194905118}{Phys. Lett. A {\bf 184}, 394 (1994)}.
\bibitem{zrsv97} S. Zapperi, P. Ray, H. E. Stanley, and A. Vespignani, \href{https://journals.aps.org/prl/abstract/10.1103/PhysRevLett.78.1408}{Phys. Rev. Lett. {\bf 78}, 1408 (1997)}.
\bibitem{zns05} S. Zapperi, P. K. V. V. Nukala, and S. Simunovic, \href{https://journals.aps.org/pre/abstract/10.1103/PhysRevE.71.026106}{Phys. Rev. E {\bf 71}, 026106 (2005)}.
\bibitem{zns05a} S. Zapperi, P. K. V. V. Nukala, and S. Simunovic, \href{https://www.sciencedirect.com/science/article/abs/pii/S0378437105005224}{Physica A {\bf 357}, 129 (2005)}.
\bibitem{dlv08} D. Dalmas, A. Lelarge, and D. Vandembroucq, \href{https://journals.aps.org/prl/abstract/10.1103/PhysRevLett.101.255501}{Phys. Rev. Lett. 101, 255501 (2008)}.
\bibitem{mm07} I. Malakhovsky and M. A. J. Michels, \href{https://journals.aps.org/prb/abstract/10.1103/PhysRevB.76.144201}{Phys. Rev. B 76, 144201 (2007)}.
\bibitem{rd96} P. Ray and G. Date, \href{https://www.sciencedirect.com/science/article/abs/pii/0378437195004319?via\%3Dihub}{Physica A (Amsterdam) 229, 26 (1996)}.
\bibitem{pmbr20} R. P. S. Parihar , D. V. Mani, A. Banerjee, and R. Rajesh, \href{}{Physical Review E {\bf 102}, 053002 (2020)}.
\bibitem{s06} V. V. Silberschmidt, \href{https://link.springer.com/article/10.1007/s10704-005-3994-8}{Int. J. Fract. 140, 73 (2006)}. 
\bibitem{szs13} A. Shekhawat, S. Zapperi, and J. P. Sethna, \href{https://journals.aps.org/prl/abstract/10.1103/PhysRevLett.110.185505}{Phys. Rev. Lett. {\bf 110}, 185505 (2013)}.
\bibitem{mohaha12} A. A. Moreira, C. L. N. Oliveira, A. Hansen, N. A. M. Araujo, H. J. Herrmann, and J. S. Andrade, Jr., \href{https://journals.aps.org/prl/abstract/10.1103/PhysRevLett.109.255701}{Phys. Rev. Lett. 109, 255701 (2012)}.
\bibitem{srvm95} J. Schmittbuhl, S. Roux, J. P. Vilotte, and K. J. Maloy, \href{https://journals.aps.org/prl/abstract/10.1103/PhysRevLett.74.1787}{Phys. Rev. Lett. 74, 1787 (1995)}.
\bibitem{ref97} S. Ramanathan, D. Ertas, and D. S. Fisher, \href{https://journals.aps.org/prl/abstract/10.1103/PhysRevLett.79.873}{Phys. Rev. Lett. 79, 873 (1997)}.
\bibitem{phss91} C. K. Peng, S. Havlin, M. Schwartz, and H. E. Stanley, \href{https://journals.aps.org/pra/abstract/10.1103/PhysRevA.44.R2239}{Phys. Rev. A 44, R2239 (1991)}.
\bibitem{xh00} Z. C. Xia and J. W. Hutchinson, \href{https://www.sciencedirect.com/science/article/abs/pii/S0022509699000812?via\%3Dihub}{J. Mech. Phys. Solids 48, 1107 (2000)}.  
\bibitem{hansenBook} A. Hansen, P. C. Hemmer, and S. Pradhan, {\em The Fibre Bundle Model}, Wiley-Vch, Germany (2015).
\bibitem{cbp21} B. K. Chakrabarti, S. Biswas, and S. Pradhan, \href{https://doi.org/10.3389/fphy.2020.613392}{Front. Phys. 8, 613392}.
\bibitem{brr15} S. Biswas, S. Roy, and P. Ray, \href{https://journals.aps.org/pre/abstract/10.1103/PhysRevE.91.050105}{Phys. Rev. E {\bf 91}, 050105(R) (2015)}. 
\bibitem{hmkh02} R. C. Hidalgo, Y. Moreno, F. Kun, and H. J. Herrmann, \href{https://journals.aps.org/pre/abstract/10.1103/PhysRevE.65.046148}{Phys. Rev. E {\bf 65}, 046148 (2002)}.
\bibitem{r17} S. Roy, \href{https://journals.aps.org/pre/abstract/10.1103/PhysRevE.96.042142}{Phys. Rev. E {\bf 96}, 042142 (2017)}.
\bibitem{srh20} S. Sinha, S. Roy, and A. Hansen, \href{https://journals.aps.org/prresearch/abstract/10.1103/PhysRevResearch.2.043108}{Phys. Rev. Research {\bf 2}, 043108 (2020)}.
\bibitem{rr15} S. Roy, and P. Ray, \href{https://iopscience.iop.org/article/10.1209/0295-5075/112/26004}{Europhysics Letters {\bf 112}, 26004 (2015)}
\bibitem{srh21} S. Sinha, S. Roy, and A. Hansen, \href{https://www.sciencedirect.com/science/article/pii/S0378437121000546}{Physica A {\bf 569}, 125782 (2021)}.  
\bibitem{rbr17} S. Roy, S. Biswas, and P. Ray, \href{https://journals.aps.org/pre/abstract/10.1103/PhysRevE.96.063003}{Phys. Rev. E {\bf 96}, 063003 (2017)}.
\bibitem{kdk17} V. Kadar, Z. Danku, and F. Kun, \href{https://doi.org/10.1103/PhysRevE.96.033001}{Phys. Rev. E 96, 033001 (2017)}.
\bibitem{kk19} V. Kadar, and F. Kun, \href{https://doi.org/10.1103/PhysRevE.100.053001}{Phys. Rev. E 100, 053001 (2019)}.
\bibitem{kpk20} V. Kadar, G. Pal, and F. Kun, \href{https://doi.org/10.1038/s41598-020-59333-4}{Scientific Reports 10, 2508 (2020)}.
\bibitem{pkh19} S. Pradhan, J. T. Kjellstadli, and A. Hansen, \href{https://doi.org/10.3389/fphy.2019.00106}{Front. Phys. 7, 106 (2019)}.
\bibitem{tmjsr19} Bosiljka Tadic, Svetislav Mijatovic, Sanja Janicevic, Djordje Spasojevic, and Geoff J. Rodgers, \href{https://doi.org/10.1038/s41598-019-42802-w}{Scientific Reports {\bf 9}, 6340 (2019)}. 
\bibitem{smpv18} Djordje Spasojevic, Svetislav Mijatovi, Victor Navas-Portella, and Eduard Vives, \href{https://doi.org/10.1103/PhysRevE.97.012109}{Phys. Rev. E {\bf 97}, 012109 (2018)}.   
\bibitem{Pierce} F. T. Pierce, \href{https://www.tandfonline.com/doi/abs/10.1080/19447027.1926.10599953}{J. Text. Inst. {\bf 17}, 355 (1926)}.
\bibitem{ft28} R. A. Fisher, and L. H. C. Tippett, \href{https://www.cambridge.org/core/journals/mathematical-proceedings-of-the-cambridge-philosophical-society/article/abs/limiting-forms-of-the-frequency-distribution-of-the-largest-or-smallest-member-of-a-sample/7BE8DE65FCDFC3ABECFE1054DFB56CB5}{Proc. Cambridge Philos. Soc. {\bf 24}, 180 (1928)}.
\bibitem{Smith} R. L. Smith and S. L. Phoenix, \href{https://asmedigitalcollection.asme.org/appliedmechanics/article-abstract/48/1/75/422789/Asymptotic-Distributions-for-the-Failure-of}{J. Appl. Mech. {\bf 48}, 75 (1981)}.
\bibitem{sgh12} A. Stormo, K. S. Gjerden, and A. Hansen, \href{https://doi.org/10.1103/PhysRevE.86.025101}{Phys. Rev. E {\bf 86}, 025101(R)(2012)}.
\bibitem{time} At a certain point during the evolution of the model, if it goes through $m$ stress increment with $n_1$, $n_2$, $\cdots$, $n_m$ redistribution steps for $1st$, $2nd$, $\cdots$, $mth$ stress increment, then the time corresponding to this scenario will be: $m + \displaystyle \sum_{m} n_m$.
\bibitem{rb21} S. Roy and S. Biswas, \href{https://doi.org/10.3389/fphy.2021.643602}{Front. Phys. {\bf 9}, 643602 (2021)}.
\bibitem{bbr21} N. K. Bodaballa, S. Biswas, S. Roy, \href{https://arxiv.org/abs/2108.06516}{ arXiv:2108.06516}.
\bibitem{bcipsssv13} J. Baro, A. Corral, X. Illa, A. Planes, E. K. H. Salje, W. Schranz, D. E. Soto-Parra, and E. Vives, \href{https://journals.aps.org/prl/abstract/10.1103/PhysRevLett.110.088702}{Phys. Rev. Lett. 110, 088702 (2013)}.
\end{thebibliography}
\end{document}